\documentclass[grl,draft]{agu2001}
\usepackage{amsmath}
\usepackage{amssymb}
\usepackage{graphicx}
\usepackage{bm}

\journalid{}
\articleid{}{}
\paperid{}
\cpright{}{}
\received{} \revised{} \accepted{} \published{}

\authorrunninghead{SWISDAK ET AL.}
\titlerunninghead{POSITIVE STORM SIMULATIONS}
\authoraddr{M. Swisdak,
Icarus Research, Inc., PO Box 30780 Bethesda MD 20824-0780, USA ,
(swisdak@ppd.nrl.navy.mil)}

\begin{document}
\title{Simulation Study of a Positive Ionospheric Storm Phase Observed
at Millstone Hill}

\authors{M. Swisdak, \altaffilmark{1}
J. D. Huba, \altaffilmark{2} G. Joyce\altaffilmark{1}, and Chao-Song
Huang\altaffilmark{3} }

\altaffiltext{1}
{Icarus Research, Inc., Bethesda, MD, USA.}

\altaffiltext{2}
{Plasma Physics Division, Naval Research Laboratory, Washington, DC, USA.}

\altaffiltext{3}
{Haystack Observatory, Massachusetts Institute of Technology,
Westford, Massachusetts}

\begin{abstract}

Simulation results from the NRL ionospheric model SAMI2 indicate that
the changes in the F-region over Millstone Hill during the geomagnetic
storm beginning on 3 April 2004 are primarily due to the influence of
a long-lasting eastward electric field, as was previously suggested by
{\it Huang et al.} [2005]\nocite{huang05a}.  A simulation of the storm
day agrees well with the observational data and shows that, compared
with the ionosphere of the previous quiet day, the peak electron
density in the F-region ($N_mF_2$) increased by a factor of
$\approx\negmedspace 2$, the altitude of the peak density ($h_mF_2$)
rose by $\approx\negmedspace 80$ km, and the F-region electron
temperature decreased by $\approx\negmedspace 1000$ K.  Further
simulations in which either the neutral atmosphere and winds or the
electric field were replaced by their quiet day counterparts clearly
suggest that the electric field played the dominant, although not
exclusive, role in producing these effects.

\end{abstract}

\begin{article}

\section{Introduction}

After an extended quiet period lasting from 29 March to 2 April 2004 a
magnetic storm began at 1412 UT on 3 April and reached a minimum Dst
of -149 nT at 0042 UT on the following day.  {\it Huang et al.} 
[2005]\nocite{huang05a} reported that this event triggered large,
long-lasting changes in the daytime ionosphere, including a strong
positive ionospheric storm phase (i.e., a period in which the F-region
electron density increased).  Understanding long-duration mid-latitude
positive storms such as this is of particular interest because they
have significant effects on large regions of the ionosphere.  {\it
Buonsanto} [1999]\nocite{buonsanto99a} noted that the generation of
such events is one of the main unresolved problems in ionospheric
research.

\begin{figure}
\noindent\includegraphics[width=3.2in]{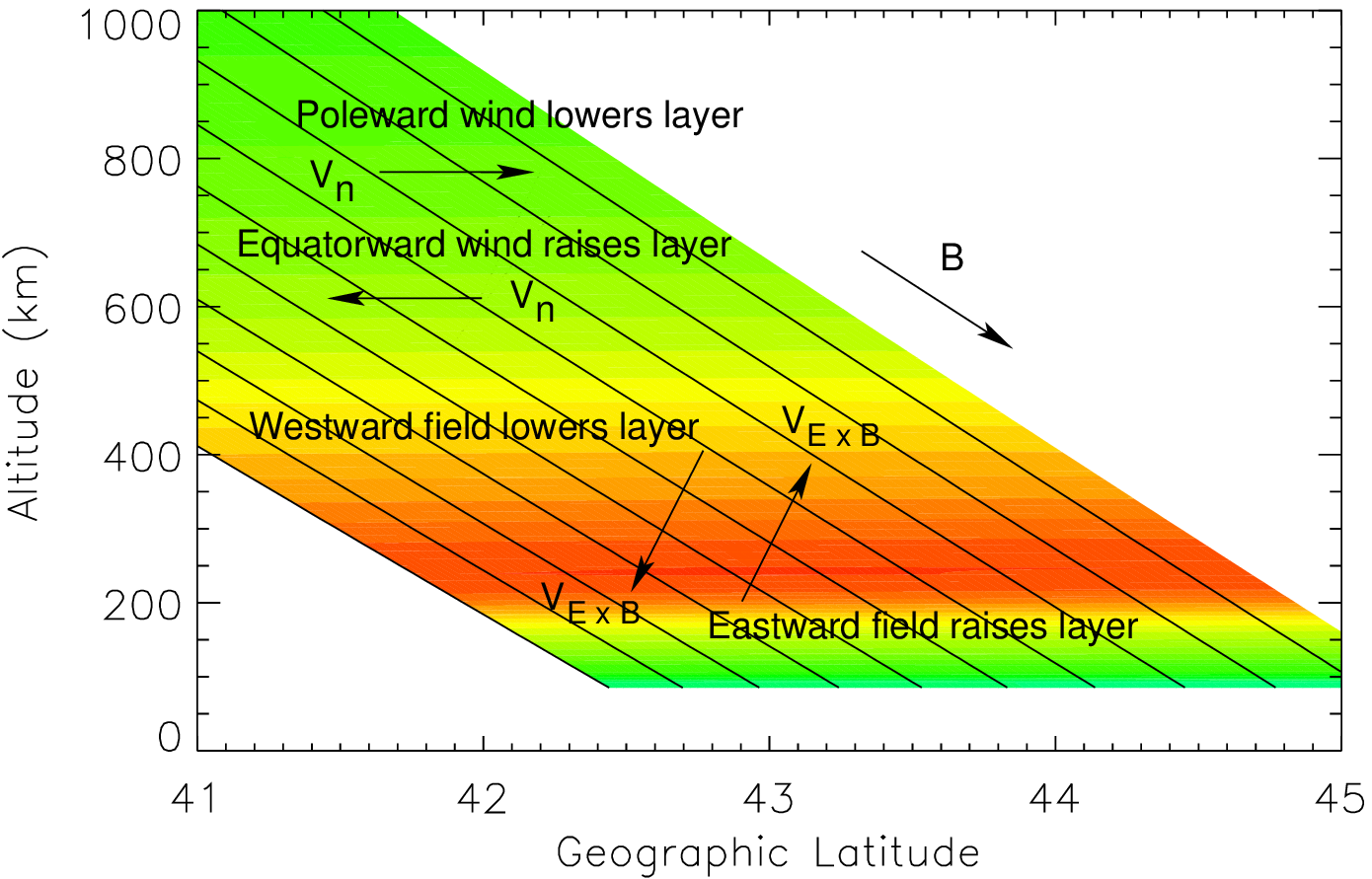}
\caption{\label{mh_cartoon}Schematic of the effects of neutral winds
and electric fields on the mid-latitude ionosphere.  The colors merely
suggest the variation of the density with altitude and do not
represent the simulations.}
\end{figure}

Two mechanisms have been proposed as drivers of such dayside storms:
winds in the neutral atmosphere and electric fields (see Figure
\ref{mh_cartoon}).  In the former, heat inputs in the auroral regions
are thought to cause global changes in the wind circulation pattern
and thermospheric composition [{\it Rishbeth et al.}, 1985; {\it
Fuller-Rowell et al.}, 1994]\nocite{rishbeth85a,fullerrowell94a},
including the generation of equatorward neutral winds that lift the
mid-latitude F-region [{\it Jakowski et al.}, 1990; {\it Bauske and
Pr\"olls}, 1997; {\it Lu et al.},
2001]\nocite{jakowski90a,bauske97a,lu01a}.  (In Figure
\ref{mh_cartoon}, $\mathbf{V_n
\cdot B} < 0$ and the collisionally-coupled plasma is driven up the
field lines.)  On the other hand, {\it Foster and Rich}
[1998]\nocite{foster98a} reported direct observations of the uplift of
the mid-latitude ionosphere by a prompt penetration eastward electric
field (in Figure \ref{mh_cartoon} the $\mathbf{E \times B}$ drift is
upward and poleward).  These processes are not completely distinct
since, for example, equatorward neutral winds can maintain a dynamo
electric field.

{\it Huang et al.} [2005]\nocite{huang05a} suggested that the
proximate cause for the 3 April storm was an enhanced eastward
electric field that lifted the mid-latitude ionosphere for several
hours.  In this Letter we present simulations that support this
conclusion, although the neutral winds do seem to play a minor, but
important, role.  We describe our computational model SAMI2 in section
\ref{model}, present the simulation results in section \ref{results},
and discuss our conclusions in section \ref{summary}.

\section{\label{model}Computational Model}

SAMI2 is a two-dimensional, semi-implicit, Eulerian fluid model of the
low to mid-latitude ionosphere at one geomagnetic longitude [{\it Huba
et al.}, 2000]. \nocite{huba00a} Previous studies have shown that
SAMI2 simulations of the F-region electron density are in good
agreement with data from both satellites [{\it Huba et al.}, 2002]
\nocite{huba02a} and the Millstone Hill observatory [{\it Huba et
al.}, 2003]. \nocite{huba03a}

In this study the simulation domain passes through a point 330 km
above Millstone Hill ($42.6^{\circ}$ N, $288.5^{\circ}$ E, invariant
latitude $55^{\circ}$) and has north-south extrema at geographic
latitudes of $-68.7^{\circ}$ and $45.2^{\circ}$.  We place 201
gridpoints along each of 114 field lines with non-uniform spacing in
both dimensions to achieve better resolution at low altitudes.  Test
runs in which the number of points in either dimension is doubled
suggest our results have converged.  The first and last gridpoints of
each field line are at an altitude of 85 km and the apexes range
between 150 and 14,000 km ($L$ values of 1.02 to 3.20).

We model the terrestrial magnetic field as an offset, tilted dipole
for which the center as well as the geographic latitude and longitude
of the magnetic north pole have been chosen to maximize agreement in
the simulation domain with the International Geomagnetic Reference
Field.  Parallel to the field, i.e., along a flux tube, we solve the
fluid continuity and momentum equations for seven ion species (H$^+$,
He$^+$, N$^+$, O$^+$, N$_2^+$, NO$^+$, and O$_2^+$) and the
temperature equation for three (H$^+$, He$^+$, and O$^+$).  The
temperatures of the other four ions are taken to be equal to that of
O$^+$.  To model the electrons we assume that the charge density and
parallel current density vanish, which then determines the electron
density and velocity parallel to the field; the electron temperature
equation is solved separately.  The empirical models NRLMSISE-00 and
HWM93 [{\it Picone et al.}, 2002; {\it Hedin et al.},
1991]\nocite{picone02a,hedin91a} specify the composition and winds of
the neutral atmosphere, respectively.

We assume that transport perpendicular to the magnetic field is solely
due to $\mathbf{E\times B}$ drifts.  To find the electric field
throughout the simulation domain we extrapolate from measurements of
the east-west ($E_x$) and north-south ($E_y$) components of the
electric field in the F-region above Millstone Hill by making two
assumptions.  The first, that every (dipolar) field line is an
equipotential, allows us to calculate the $\mathbf{E\times B}$ drift
everywhere on a field line once we know it anywhere on a field line.
The second, that the (vertical) drift varies as $L^2$ at the magnetic
equator, is valid when the electric field is curl-free and the
azimuthal neutral wind dynamo is negligible [{\it Eccles},
1998]\nocite{eccles98a}.  Given a drift at Millstone Hill of magnitude
$v_{M}$ the magnitude of the drift at any other point in the
simulation domain is
\begin{equation}\label{exb}
v_{E\times B} = v_{M}\sqrt{\frac{1+3
\cos^2\theta_{M}}{1+3\cos^2\theta}}
\left(\frac{\sin\theta}{\sin\theta_{M}}\right)^3
\left(\frac{L}{L_{M}}\right)^2 \mbox{,}
\end{equation}
where $\theta$ is the magnetic co-latitude.  The direction of the
drift is always perpendicular to $\mathbf{B}$ and hence varies along a
field line. 

The incoherent scatter radar at Millstone Hill measures
three-dimensional ion velocities (from which the electric field can be
determined) and meridional neutral winds with a time resolution of
$\approx\negmedspace 30$ minutes.  Typical errors are $\pm 0.2$ mV/m
for the electric field and $\pm 20$ m/s for the wind.  {\it Huang et
al.} [2005] \nocite{huang05a} give a more complete description of the
measurements.  The two relevant components of the electric field as
well as the corresponding $\mathbf{E \times B}$ drifts are plotted in
the top two panels of Figure \ref{ez_win}.  Because of the non-zero
magnetic declination at Millstone Hill ($\approx\negmedspace
-15^{\circ}$) both the east-west and north-south components contribute
to the drift, although the north-south contribution is minimal until
$\approx\negmedspace 20$ UT on 3 April.  At Millstone Hill a 3 mV/m
east-west electric field implies a total drift speed of
$\approx\negmedspace 60$ m/s and a vertically projected drift speed of
$\approx\negmedspace 20$ m/s.

Unfortunately, although we have measurements of the average F-region
meridional neutral wind above Millstone Hill, there is no
straightforward way to extrapolate this data to the entire simulation
domain.  Moreover, we have no measurements of the zonal winds.  We
instead use the velocities (both meridional and zonal) from the
empirical model HWM93.  The model values for the meridional wind are
plotted in the bottom panel of Figure \ref{ez_win} along with the
observations.  While the basic features of the data sets agree there
are some notable differences, particularly in the magnitude of the
wind during the early evening and pre-dawn hours (local time at
Millstone Hill = UT-5).  Possible effects of these discrepancies are
discussed further in Section \ref{summary}.  Note that because of the
large dip angle at Millstone Hill only 1/3 of the meridional wind
speed is projected along the magnetic field.

SAMI2's empirical models of the neutrals and solar flux depend on the
geophysical parameters F$_{10.7}$, F$_{10.7\text{A}}$ and Ap --- the
previous day's solar flux at 10.7 cm, the 81-day centered average of
F$_{10.7}$, and the Ap index.  For both days F$_{10.7\text{A}}=
105.1$; on 2 April F$_{10.7}= 114.1$ and on 3 April F$_{10.7}= 108.6$.
The daily $\text{Ap}= 3$ on 2 April and $41$ on 3 April, but we also
used finer gradations (e.g., 3-hour ap indices) in NRLMSISE-00 and
HWM93.

\section{\label{results}Simulation Results}

\begin{figure}
\noindent\includegraphics[width=3.2in]{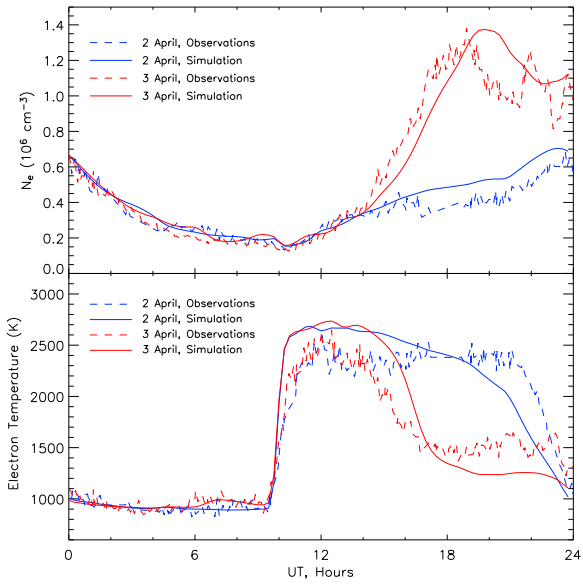}
\caption{\label{ne} Comparison between observations (dashed lines) and
simulation (solid) of the electron density (top) and temperature
(bottom) at 330 km above Millstone Hill.  Values from 2 April are
shown in blue, 3 April in red.}
\end{figure}

In Figure \ref{ne} we compare our simulation results to the observed
electron densities and temperatures at an altitude of 330 km above
Millstone Hill.  The overall agreement is good, with the simulation
successfully modeling the major changes between the quiet and active
days.  After the storm begins ($\approx\negmedspace 14$ UT on 3 April)
the electron density quickly increases, peaking at
$\approx\negmedspace 1.4 \times 10^6\text{ cm}^{-3}$, or roughly $2-3$
times the density at the same time on the previous day.
Simultaneously the electron temperature drops by $\approx\negmedspace
1000$ K.

However several discrepancies can be seen.  The first is in the
electron temperature on 2 April when the model overshoots the observed
value by roughly 10\% at 12 UT and remains too high for several
hours. This is probably due to the photoelectron heating model used in
SAMI2 and is discussed further in Section \ref{summary}. A second
discrepancy is the $\sim\negmedspace$ 1 hour lag between the onset of
the simulated and actual storm on 3 April.  Winds can cause such a
delay by retarding the flow of material up a field line, but both the
modeled and true meridional winds are relatively modest at this time.
However HWM93 predicts a relatively large zonal wind ($\simeq$ 100
m/s) during this period which, when projected onto the magnetic field,
is large enough to cause the delay.  An otherwise identical simulation
that was performed with no zonal winds exhibited no lag in the storm
onset.

To better show the effects of the storm on the F-region as a whole we
plot {\it N$_\text{m}$F$_2$} and {\it h$_\text{m}$F$_2$} at Millstone
Hill for both the observations and the simulation in Figure
\ref{tec}. $h_mF_2$ remained below 300 km during daylight hours on 2
April but during the storm on the following day it rose by $50-80$ km.
Viewed from a fixed altitude of 330 km the rise of the F-region leads
to an increase in the local electron density and a decrease in the
electron temperature (see the top panel of Figure \ref{ne}), i.e.,
cooler, denser plasma moves to higher altitudes.  Note that this
mechanism does not depend on what process lifts the F-region.

To test the relative importance of the electric field and the neutral
atmosphere and winds in driving these large changes we performed two
further simulations of the storm day.  In the first we replaced the
models of the neutral atmosphere and winds with their quiet day
counterparts; in the second we used the storm day neutral atmosphere
and winds and the quiet day electric field.  The results are shown in
Figure \ref{lowe}.

For the storm day electric field and the quiet day neutrals (dotted
black line) the largest change from the original simulation is the
$\sim 2$ hour delay in the ionosphere's response to the storm's onset.
We attribute this to the neutral wind. The quiet day neutral wind is
poleward which, as noted earlier, pushes plasma down the field line
and suppresses, albeit not completely, the increase in electron
density after 14 UT.  There are also some minor differences in the
temporal evolution of the electron density, particularly late in the
storm ($20-22$ UT) when the simulation density rises as the observed
density falls.  In comparison the simulation with the quiet field and
storm neutrals correctly captures the onset of the storm but diverges
from the observations after 16 UT. In particular, this simulation
underestimates the peak density by $\simeq 50\%$. We attribute this to
the quiet day electric field that changes from eastward to westward at
$\simeq$ 18 UT, thus pushing the F layer downward and reducing the
electron density.  Together these simulations suggest that the
neutrals played a role in the initial stages of the storm but the
electric field was the principal driver of the ionospheric evolution.

\section{\label{summary}Discussion}

\begin{figure}
\noindent\includegraphics[width=3.2in]{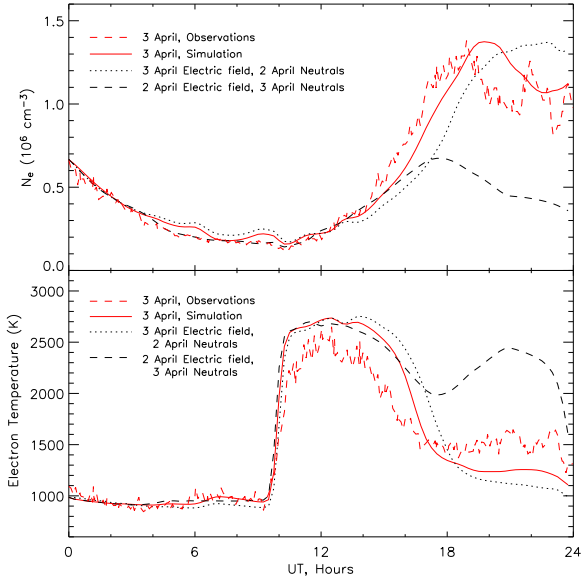}
\caption{\label{lowe} Electron density (top) and
temperature (bottom) for simulations of 3 April.  The observational
data (dashed red) and normal simulation (solid red) are the same as in
Figure \ref{ne}.  For the dashed black line the simulated electric
field of 3 April has been replaced with that of 2 April.  For the
dotted black line the electric field is normal but the neutral
atmosphere and winds have their 2 April values.}
\end{figure}

We have presented a simulation study of storm-time effects on the
mid-latitude ionosphere over Millstone Hill observatory using the NRL
ionosphere code SAMI2.  The simulation results agree reasonably well
with the observations on both the quiet and the storm days.  In
particular, the model predicts the changes in the ionosphere over
Millstone Hill relative to the previous (quiet) day: the F-peak
altitude rose by $\approx\negmedspace 80$ km, the F-peak electron
density increased by a factor of $\approx\negmedspace 2$, and the
F-region electron temperature decreased by $\approx\negmedspace 1000$
K. We primarily attribute these dramatic changes to the long-lasting
eastward electric field observed on the storm day between 12--20 UT
that lifts cold, dense plasma to higher altitudes.  We base this
conclusion in large part on Figure \ref{lowe}, which indicates that
the storm day neutral wind and quiet day electric field do not
sufficiently account for the observations. By contrast, the active day
electric field and quiet day neutral wind do capture the salient
effects of the storm: a large enhancement in the electron density and
decrease in the electron temperature.

\begin{figure}
\noindent\includegraphics[width=3.2in]{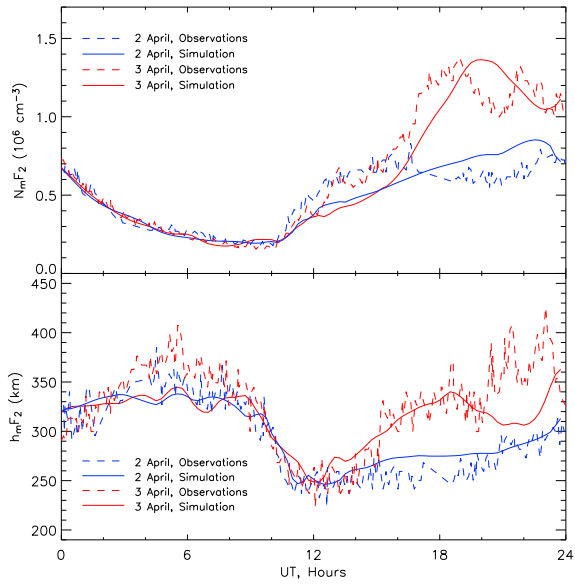}
\caption{\label{tec} Comparison between observations (dashed lines)
and simulation (solid) of {\it N$_\text{m}$F$_2$} (top) and {\it
h$_\text{m}$F$_2$} (bottom) above Millstone Hill.  Values from 2 April
are shown in blue, 3 April in red.}
\end{figure}

The variations of the electron density and temperature with altitude
were also measured around 19 UT on both the quiet and active days.
Not surprisingly, in light of the data shown in Figure \ref{tec}, {\it
Huang et al.} [2005]\nocite{huang05a} found that the storm day had
higher electron densities, lower electron temperatures, and higher
$h_mF_2$s.  Although the simulation agrees with these trends there are
discrepancies, particularly in the temperature, at higher altitudes.

These differences may be due to SAMI2's treatment of photoelectron
heating.  Collisions are sufficiently frequent at low altitudes (below
roughly 250 km, although there is some variation with the neutral
density) that we assume photoelectrons deposit their energy locally.
Above that point our model expresses the (non-local) heating as a
function of the integral of the electron density along a field line.
More sophisticated, but computationally intensive, approaches
discretize the electron distribution function in energy space and
solve some form of a Boltzmann transport equation.  The transition
between local and non-local heating is continuous in SAMI2, but during
non-equilibrium periods unphysical short-lived temperature plateaus
occasionally develop around 450 km.  By making ad hoc adjustments to
the details of our model we have established that these features have
only minimal effects on the plasma at lower altitudes.

\begin{figure}
\noindent\includegraphics[width=3.2in]{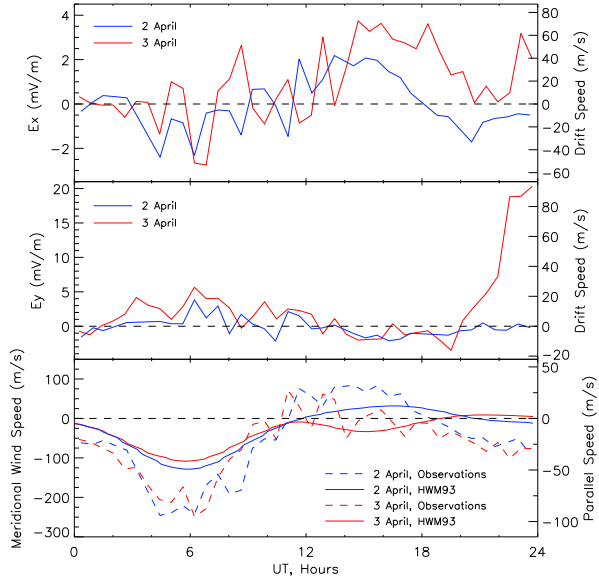}
\caption{\label{ez_win} F-region electric field components and neutral
winds above Millstone Hill on 2 April (blue) and 3 April (red). For
both the east-west ($E_x$, top panel) and north-south ($E_y$, middle
panel) components the lines denote the observations and the
simulations.  The right axis gives the magnitude of the
$\mathbf{E\times B}$ drift at 330 km due to each component.  In both
panels upward and poleward drifts are positive.  For the meridional
neutral winds (bottom panel) the solid lines show the velocities from
HWM93 and the dashed lines show the observations.  The right axis
gives the projection of the wind speed onto the local magnetic field.
Positive values are northward.}
\end{figure}

A thorough study of the origin of the prolonged, storm-time eastward
electric field would require a coupled ionosphere-magnetosphere model
that is beyond the scope of this Letter.  However two possible sources
are (1) a penetration electric field associated with a rapid change in
the inner magnetospheric electric field at storm onset (e.g., {\it
Kikuchi and Araki} [1979]\nocite{kikuchi79a}, {\it Foster and Rich}
[1998]\nocite{foster98a}) and (2) a wind driven dynamo field
associated with high-latitude heating of the atmosphere and the
generation of equatorward neutral winds (e.g., {\it Fuller-Rowell et
al.} [1994])\nocite{fullerrowell94a}. {\it Huang et al.} [2005] argued
for a penetration electric field because it can quickly propagate to
low latitudes, in agreement with the minimal lag between the storm
onset and the ionospheric response seen in the data.  Furthermore, no
strong equatorward winds were observed at Millstone Hill during the
storm. (see the bottom panel of Figure \ref{ez_win}).

A better comparison between simulations and data could be made with
more realistic values for the neutral winds throughout the E and F
regions.  These can be obtained through either more detailed
observations or a coupled thermosphere-ionosphere model.  In the short
term we will pursue the former approach using, for example,
measurements taken at Millstone Hill for the September 2005 ISR World
Month campaign (L. P. Goncharenko, private communication).  Finally,
we will also use SAMI3, an extension of SAMI2 to all longitudes, to
investigate longitudinal effects such as the possible generation of
inhomogeneous total electron content (TEC) enhancements by a poleward
electric field ({\it Vlasov et al.} [2003]).\nocite{vlasov03a}

\begin{acknowledgments}
This work was supported by the Office of Naval Research.  Work at MIT
Haystack Observatory was supported by an NSF cooperative agreement
with the Massachusetts Institute of Technology.
\end{acknowledgments}

\end{article}

\end{document}